\documentclass{aa}
\newcommand {\hii}{H\,{\sc ii}} 
\newcommand {\hei}{He\,{\sc i}} 
\newcommand {\heii}{He\,{\sc ii}} 
\newcommand {\kms}{\relax \ifmmode {\,\rm km\,s}^{-1}\else \,km~s$^{-1}$\fi}
\newcommand {\ha}{H$\alpha$}
\newcommand {\hb}{H$\beta$}
\newcommand {\hg}{H$\gamma$}
\newcommand {\hd}{H$\delta$}
\newcommand {\he}{H$\epsilon$}
\newcommand {\oiii}{[O\,{\sc iii}]}
\newcommand {\oii}{[O\,{\sc ii}]}
\newcommand {\oi}{[O\,{\sc i}]}
\newcommand {\no}{[N\,{\sc i}]}
\newcommand {\nii}{[N\,{\sc ii}]}
\newcommand {\sii}{[S\,{\sc ii}]}
\newcommand {\siii}{[S\,{\sc iii}]}
\newcommand {\feiii}{[Fe\,{\sc iii}]}
\newcommand {\feii}{[Fe\,{\sc ii}]}
\newcommand {\cliii}{[Cl\,{\sc iii}]}
\newcommand {\neiii}{[Ne\,{\sc iii}]}
\newcommand {\neiv}{[Ne\,{\sc iv}]}
\newcommand {\nev}{[Ne\,{\sc v}]}
\newcommand {\ariv}{[Ar\,{\sc iv}]}

\usepackage{graphicx}
\usepackage{txfonts}
\usepackage{rotating}

\begin{document}

   \title{WR bubbles and \heii\ emission\thanks{Based on observations 
collected at the European Southern Observatory, Cerro Paranal, Chile 
(ESO No. 68.C-0238(A,B)).}}

   \subtitle{}

   \author{Y. Naz\'e\thanks{Research Fellow FNRS (Belgium)}\inst{1}
	  G. Rauw\thanks{Research Associate FNRS (Belgium)} \inst{1}
          J. Manfroid\thanks{Research Director FNRS (Belgium)}  \inst{1}
	  Y.-H. Chu \inst{2} \and J.-M. Vreux \inst{1}}

   \offprints{Y. Naz\'e \email{naze@astro.ulg.ac.be}}

   \institute{Institut d'Astrophysique et de G\'eophysique;
	Universit\'e de Li\`ege;
	All\'ee du 6 Ao\^ut 17, Bat. B5c;
	B 4000 - Li\`ege;
	Belgium 
\and Astronomy Department;
University of Illinois at Urbana-Champaign;
	1002 West Green Street;
	 Urbana, IL 61801;
	USA
             }
\authorrunning{Naz\'e et al.}
   \date{}

   \abstract{

We present the very first high quality images of the \heii\
$\lambda$4686 emission in three high excitation nebulae of the 
Magellanic Clouds. A fourth high excitation nebula, situated 
around the WR star BAT99-2, was analysed in a previous letter.
Using VLT FORS data, we investigate the morphology of the ring nebulae 
around the early-type WN stars BAT99-49 \& AB7.  We 
derive the total \heii\ fluxes for each object and compare them 
with the most recent theoretical WR models. Whilst the ionization
of the nebula around BAT99-49 can be explained by a WN star of 
temperature 90-100~kK, we find that the \heii\ emission measure of
the nebula associated with AB7 requires an He$^+$ ionizing flux
larger than predicted for the hottest WN model available. Using \ha, 
\oiii\ and \hei\ $\lambda$5876 images along with long-slit spectroscopy,
we investigate the physical properties of these ring nebulae and find
only moderate chemical enrichment. 

We also surveyed seven other LMC WR stars but we failed to detect any 
\heii\ emission. This holds also true for BAT99-9 which had been proposed 
to excite an \heii\ nebula. Four of these surveyed stars are surrounded 
by a ring nebula, and we use the FORS data to derive their chemical 
composition: the nebula around BAT99-11 shows a N/O ratio and an
oxygen abundance slightly lower than the LMC values, 
while the nebula around BAT99-134 presents moderate 
chemical enrichment similar to the one seen near BAT99-2, 49 and AB7. 
Comparing the WR stars of the LMC, BAT99-2 and 49 appear 
rather unique since similar stars do not reveal high excitation 
features.

The third high excitation nebula presented in this paper, N44C, does not harbor 
stars hotter than mid-O main sequence stars. It was suggested to be 
a fossil X-ray nebula ionized by the transient LMC X-5.
Our observations of N44C reveal no substantial
changes in the excitation compared to previous results reported
in the literature. Therefore, we conclude that either the
recombination timescale of the X-ray nebula has been underestimated 
or that the excitation of the nebula is produced by 
another, yet unknown, mechanism.

   }

   \maketitle
\section{Introduction}

\begin{table*} 
\begin{center}
\caption{Total exposure times (in sec.) for each object, in each filter 
and for each grism. Spectral types are taken from Foellmi et al. 
(\cite{foesmc} and \cite{foelmc}), Bartzakos et al. \cite{bar} and 
Niemela et al. \cite{nie02}.\label{texp}} 
\begin{tabular}{l | c | c c c c c c | c c} 
\hline\hline
Object& Sp. Type&\multicolumn{6}{c|}{Imaging Filters}& \multicolumn{2}{c}{Grisms}\\
& &\ha\ & \oiii\ & \hei$\lambda$5876 & \heii$\lambda$4686& cont. 5300& cont. 6665& 600B& 600V\\ 
\hline
LMC BAT99-2$^a$ & WN2b(h)&300& 300& 3600& 3600& 540& 300& 900& 900\\
LMC BAT99-49 & WN4:b+O8V&300& 300& 2700& 2700& 405& 300& 900& 900\\
SMC AB7 & WN2-4+O6If&300& 300& 2400& 2800& 240& 300& 900& 900\\
LMC N44C & &300& 300& 1350& 1400& 198& 300& 500& 600\\
\hline
LMC BAT99-8 & WC4&& & & & & & 300& 100\\
LMC BAT99-9 & WC4&& & & & & & 300& \\
LMC BAT99-11 & WC4&& & & & & & 300& 300\\
LMC BAT99-52 & WC4&& & & & & & 300& \\
LMC BAT99-63 & WN4ha:&& & & & & & 300& 200\\
LMC BAT99-84 & WC4&& & & & & & 300& \\
LMC BAT99-134 & WN4b&& & & & & & 300& 300\\
\hline
\end{tabular}
\end{center}
 $^a$ See Paper I.
\end{table*}

\begin{table*} 
\begin{center}
\caption{Dereddened line ratios with respect to \hb=100.0 (see text 
for details). Uncertainties in the line ratios, estimated from the 
signal/noise in the lines and the calibration errors, are given in 
parentheses. The physical properties derived using the line ratios 
are also indicated below. A `:' denotes an uncertain value, that is 
not used for abundance calculation when other measurements of the 
same ion exist.
\label{lineratwr}} 
\begin{tabular}{l | c c c | c c c c c } 
\hline\hline
& \multicolumn{3}{c|}{LMC BAT99-49}& \multicolumn{5}{c}{SMC AB7}\\
 & N& S1 & S2 & W1 & W2 & E1 & E2 & E3 \\ 
\hline
\oii\ 3727& 178 (19) & 244 (25) & 262 (27) & 50 (5) & 121 (13) & 82 (9) & 100 (11) & 251 (27) \\
H12& & 9.5: & 5.0 (0.5) & 6.6 (0.7) & & 7.7 (0.8) & 6.5 (0.7) & 5.4 (0.6) \\
H11&  & 12: & 6.9 (0.7) &  &  & 6.3 (0.7) & 5. (0.5) & 5.3 (0.6) \\
H10& & 10: & 6.9 (0.7) & 9.4 (1.) & 8.1 (0.1) & 9.6 (1.) & 7.9 (0.8) & 8.2 (0.9) \\
H9& & 12: & 9.1 (0.9) & 8.8 (0.9) & 8.5 (0.1) & 9.2 (1.) & 8.5 (0.9) & 7.9 (0.8)\\
\neiii\ 3868&  41 (4) & 44 (4) & 57 (6) & 55 (6) & 55 (6) & 47 (5) & 57 (6) & 59 (6)\\
H8 + \hei& 18 (2) & 20 (2) & 20 (2) & 14 (1) & 16 (2) & 17 (2) & 18 (2) & 20 (2) \\
\neiii\ + \he & 26 (2) & 27 (3) & 33 (3) & 31 (3) & 29 (3) & 30 (3) & 34 (3) & 35 (3)\\
\hd &  28 (2) & 29 (3) & 28 (3) & 30 (3) & 29 (3) & 31 (3) & 30 (3) & 30 (3) \\
\hg &  55 (4) & 51 (4) & 50 (4) & 49 (4) & 49 (4) & 50 (4) & 50 (4) & 50 (4)\\
\oiii\ 4363& 5.8 (0.5) & 5. (0.4) & 4.9 (0.4) & 14 (1) & 8.9 (0.7) & 11 (1) & 13 (1) & 9.4 (0.8) \\
\hei\ 4471& & 2.6: & 4.4 (0.3) & 3. (0.2) & 3.7 (0.3) & 2.7 (0.2) & 3.2 (0.2) & 4.2 (0.3) \\
\heii\ 4686& 8.1 (0.6) & 9.8 (0.7) &  & 24 (2) & & 21 (1) & 16 (1) \\
\ariv\ 4711$^a$& & & & 4. (0.3) & & 3.4 (0.3) & 2.9 (0.2) \\
\ariv\ 4740& & & & 3.2 (0.2) & & 2.6 (0.2) & 2.4 (0.2) \\
\hb & 100. & 100. & 100.& 100.& 100.& 100. & 100. & 100. \\
\oiii\ 4959& 157 (11) & 155 (11) & 191 (14) & 224 (16) & 200 (14) & 194 (14) & 229 (16) & 194 (14) \\
\oiii\ 5007& 471 (34) & 464 (33) & 566 (40) & 670 (48) & 597 (43) & 583 (42) & 685 (49) & 573 (41) \\
\heii\ 5412& & & & 1.7 (0.1) & & 1.6 (0.1) & 1.2 (0.1) &  \\
\hei\ 5876 & 12 (1) & 12 (1) & 13 (1) & 8.5 (0.7) & 11 (1) & 8.4 (0.7) & 9. (0.1) & 11 (1) \\ 
\oi\ 6300  & & 4.4 (0.4) & 7.1 (0.7) & & 3.5 (0.3) & 1.7 (0.2) & 2.7 (0.3) & 6.3 (0.6) \\
\siii\ 6312& & 1.7 (0.2) & 2.2 (0.2) & 1.8 (0.2) & 2. (0.2) & 1.7 (0.2) & 1.9 (0.2) & 2.1 (0.2) \\
\oi\ 6363& & 1.6 (0.2) & 2.2 (0.2) & & 1. (0.1) & & 0.8 (0.2) & 1.9 (0.2) \\
\nii\ 6548 & 7.2 (0.7) & 7.9 (0.8) & 8.9 (0.9) &1. (0.1) & 1.9 (0.2) & 1.6 (0.2) & 1.4 (0.1) & 3.8 (0.4) \\
\ha$^b$& 282 (28) & 282 (28) & 286 (29) & 279 (28) & 282 (28) & 279 (28) & 279 (28) & 279 (28)\\
\nii\ 6583& 16 (2) & 24 (2) & 28 (3) & 2.3 (0.2) & 5.6 (0.6) & 3.6 (0.4) & 4.2 (0.4) & 11 (1) \\
\hei\ 6678& 2.7 (0.3) & 3.3 (0.3) & 3.7 (0.4) & 2.5 (0.3) & 3. (0.3) & 2.6 (0.3) & 2.7 (0.3) & 3.1 (3) \\
\sii\ 6716& 13 (1) & 29 (3) & 36 (4) & 4.1 (0.4) & 13 (1) & 7.6 (0.8) & 9.2 (1.) & 22 (2) \\
\sii\ 6731& 9.0 (0.9) & 20 (2) & 25 (3) & 2.8 (0.3) & 9.2 (1.) & 5.4 (0.6) & 6.4 (0.7) & 16 (2) \\
$F$(\hb) (10$^{-14}$ erg cm$^{-2}$ s$^{-1}$)& 0.5& 1.4& 4.6& 5.8& 5.0& 2.3& 7.9& 5.1\\
\hline
$A_{{\rm V}}$(mag) & 0.44 & 0.44 & 0.49  & 0.29& 0.39& 0.30& 0.40& 0.47\\
$T_{{\rm e}}$\oiii\ (kK) & 12.5$\pm$0.5& 11.9$\pm$0.4 & 11.$\pm$0.4 & 15.5$\pm$0.7 & 13.5$\pm$0.5 & 14.8$\pm$0.7& 15.1$\pm$0.7 & 14.$\pm$0.6 \\
$n_{{\rm e}}$\sii\ (cm$^{-3}$) & $<$210& $<$210& $<$230& $<$160& $<$260& $<$250& $<$210& $<$230\\
$n_{{\rm e}}$\ariv\ (cm$^{-3}$) & & & & $<$3100& & $<$2700& 600-3600& \\
\hline
He$^+$/H$^+\times 10^{2}$ & 8.3$\pm$0.6& 8.9$\pm$0.6 & 9.2$\pm$0.5 & 6.6$\pm$0.3 & 8.1$\pm$0.4 & 6.5$\pm$0.3& 7.1$\pm$0.4 & 8.8$\pm$0.5 \\
He$^{2+}$/H$^+\times 10^{2}$ & 0.68$\pm$0.05& 0.82$\pm$0.06 & & 2.$\pm$0.1 & & 1.8$\pm$0.1& 1.4$\pm$0.1 & \\
$\rightarrow$ He/H$\times 10^{2}$ & 9.0$\pm$0.6& 9.8$\pm$0.8 & 9.2$\pm$0.5 & 8.6$\pm$0.1 & 8.1$\pm$0.4 & 8.3$\pm$0.4&8.5$\pm$0.4&8.8$\pm$0.5\\
O$^{0+}$/H$^+\times 10^{6}$ & & 4.9$\pm$0.3 & 9.5$\pm$0.6 & & 2.3$\pm$0.2 & & 1.3$\pm$0.2 & 3.8$\pm$0.3 \\
O$^+$/H$^+\times 10^{5}$ & 2.8$\pm$0.3& 4.5$\pm$0.5 & 6.6$\pm$0.7 & 0.39$\pm$0.04 & 1.4$\pm$0.2 & 0.72$\pm$0.08& 0.83$\pm$0.09 & 2.7$\pm$0.3 \\
O$^{2+}$/H$^+\times 10^{5}$ & 8.1$\pm$0.3& 9.3$\pm$0.4 & 14.5$\pm$0.6 & 6.5$\pm$0.3 & 8.3$\pm$0.4 & 6.4$\pm$0.3& 7.1$\pm$0.3 & 7.3$\pm$0.3\\
$\rightarrow$ O/H$\times 10^{4}$ & 1.2$\pm$0.1& 1.6$\pm$0.1 & 2.2$\pm$0.1 & 0.89$\pm$0.04 & 1.$\pm$0.04 & 0.91$\pm$0.04& 0.96$\pm$0.04 & 1.$\pm$0.1\\
N$^+$/H$^+\times 10^{6}$ & 2.1$\pm$0.2& 3.$\pm$0.2 & 4.2$\pm$0.3 & 1.9$\pm$0.1 & 5.4$\pm$0.4 & 3.3$\pm$0.2& 3.2$\pm$0.2 & 9.9$\pm$0.7 \\
($\rightarrow$ N/O$\times 10^{2}$)$^c$ & 7.7$\pm$1.& 6.6$\pm$0.8 & 6.3$\pm$0.8 & 5.$\pm$0.6 & 3.8$\pm$0.5 & 4.6$\pm$0.6& 3.9$\pm$0.5 & 3.7$\pm$0.5 \\
S$^+$/H$^+\times 10^{7}$ & 3.1$\pm$0.2& 7.6$\pm$0.6 & 11.2$\pm$0.8 & 0.64$\pm$0.05 & 2.7$\pm$0.2 & 1.3$\pm$0.1& 1.5$\pm$0.1 & 4.2$\pm$0.2 \\
S$^{2+}$/H$^+\times 10^{6}$ & & 2.$\pm$0.2 & 3.6$\pm$0.3 & 0.92$\pm$0.09 & 1.5$\pm$0.1 & 0.96$\pm$0.09& 1.$\pm$0.1 & 1.4$\pm$0.1\\
Ar$^{3+}$/H$^+\times 10^{7}$ & & & & 2.2$\pm$0.1 & & 2.$\pm$0.1& 1.7$\pm$0.1 & \\
Ne$^{2+}$/H$^+\times 10^{5}$ & 2.$\pm$0.2& 2.5$\pm$0.2 & 4.3$\pm$0.4 & 1.4$\pm$0.1 & 2.1$\pm$0.2 & 1.3$\pm$0.1& 1.5$\pm$0.2 & 2.$\pm$0.2\\
\hline
\end{tabular}
\end{center}
$^a$ The small contamination due to \hei\ 4713 was corrected using the  
strength of \hei\ 5876 and the theoretical \hei\ ratios from Benjamin 
et al. (\cite{ben}).\\
$^b$ Contamination due to \heii\ 6560 negligible.\\
$^c$ Assuming $N({\rm N}^+)/N({\rm O}^+)=N({\rm N})/N({\rm O})$.\\
\end{table*}

Massive stars are known to possess strong stellar winds. Throughout
their lives, these winds are able to shape their environment.
Winds of O stars can sweep up the Interstellar Medium (ISM) and
create interstellar bubbles, a few examples of which have been given in
Naz\'e et al. (\cite{naz01}, \cite{naz02}). 
While evolving into a Luminous Blue Variable or a Red Supergiant phase,
the star blows a very dense and slow wind that can be swept up by the
subsequent faster Wolf-Rayet (WR) wind. This second wind-blown bubble will
ultimately collide with the first one. Numerical simulations of such 
processes have been performed by e.g. Garc\'{\i}a-Segura et al. 
(\cite{gar}, and references therein). 

Many examples of ring nebulae around WR stars are known, both in the
Galaxy and in the Magellanic Clouds (MCs). However, the exact nature
and evolutionary status of these candidate bubbles are still unknown
in most cases, but could be investigated by means of kinematic studies
(e.g. Chu et al. \cite{chu99}) and/or abundance analyses.

A few WR ring nebulae present an additional, peculiar
feature: nebular \heii\ $\lambda$4686 emission. Such a high excitation 
feature is common amongst Planetary Nebulae, but O and WR stars were 
often thought unable to ionize He$^+$ in a detectable manner in their 
surroundings. Only 7 \heii\ nebulae are known in the Local Group
(Garnett et al. \cite{gar91a}), and five of them are around WR stars. 
Nebular \heii\ features have also been found farther away, in starburst 
galaxies, and were again attributed to WR stars (Schaerer \cite{sch}), 
though this interpretation is now questioned by Smith 
et al. (\cite{smi}). Observations of the \heii\ emission can provide 
an accurate constraint on the otherwise unobservable far UV 
fluxes of WR stars, but for the 
moment, this can only be done with precision for the Local Group objects. 

   \begin{figure*} 
   \centering
   \caption{FORS \ha\ image of LMC BAT99-49. 
The different regions used for spectral analysis 
are marked by a solid line, except for the background region, 
which is just outside the image. Bright stars and features 
discussed in the text are labelled.}
              \label{br40aha}
    \end{figure*}

   \begin{figure*} 
   \centering
   \caption{FORS \ha\ image of SMC N76. 
The different regions used for spectral analysis 
are marked by a solid line, and bright stars and
features discussed in the text are labelled.}
              \label{ab7ha}
    \end{figure*}

We thus decided to investigate the case of the three high excitation nebulae 
around WR stars in the MCs. The nebula around BAT99-2 was already analysed in Naz\'e et al. 
(\cite{naz03}, hereafter Paper I) and we discuss here the two remaining cases. 
To date, only low resolution, low signal/noise \heii\ images of the nebulae 
surrounding BAT99-2 and AB7 exist (Melnick \& Heydari-Malayeri 
\cite{mel}) and we present here the very first high quality VLT
images of the high excitation nebulae associated with BAT99-49 and AB7. 
In addition, we will also investigate a fourth peculiar 
\heii\ nebula, N44C, which may be a fossil X-ray nebula. 
All these objects are situated in the LMC, except for AB7, which is in the SMC.
Finally, we also present the results of a small
survey of LMC WR stars to search for the possible existence of 
nebular \heii\ emission. Since some of these stars have ring nebulae
around them, our spectroscopic data also enable us to examine 
their physical properties.
In this paper, we will first describe the observations, then explore
the morphology and spectrophotometry of the high excitation nebulae and
present our small WR survey. We will finally conclude in Sect. 5.

\section{Observations}

We obtained CCD images of the \heii\ emission nebulae in the MCs 
with the FORS1 instrument on the 8~m VLT-UT3 in 2002 January. 
The images were taken through several filters: \ha\ ($\lambda_0$=6563\AA, 
FWHM=61\AA), \oiii\ ($\lambda_0$=5001\AA, FWHM=57\AA), 
\hei\ $\lambda$5876 ($\lambda_0$=5866\AA, FWHM=60\AA), 
\heii\ $\lambda$4686 ($\lambda_0$=4684\AA, FWHM=66\AA), 
plus continuum filters centered at 5300\AA\ (FWHM=250\AA)
and 6665 \AA\ (FWHM=65\AA). 
To avoid saturation, we split the observations
into multiple, short exposures: the total exposure times achieved
are presented in Table \ref{texp}. The pixel size of the detector is
0\farcs2 on the sky and the seeing was $\sim$1\arcsec, thus providing 
the highest resolution images so far of these nebulae in this filter set. 
The data were reduced with {\sc iraf}\footnote{{\sc IRAF} is distributed 
by the National Optical Astronomy Observatories, which are 
operated by the Association of Universities for Research
 in Astronomy, Inc., under cooperative agreement with the National
 Science Foundation.} using standard methods for 
overscan and bias subtraction and flatfielding. The images of each 
filter were then aligned and combined with {\sc iraf}. We constructed 
star-free images in the \ha\ and \oiii\ lines by subtracting the 
images obtained in the neighbouring continuum filters. However, 
these continuum filters are slightly polluted by faint nebular
lines. While these faint lines constitute only small contamination compared 
to the bright \ha\ and \oiii\ lines, they are more problematic 
when we are dealing with the images corresponding to the filters centered on 
the fainter \hei\ and \heii\ lines. To obtain star-free \hei\ and
\heii\ images, we first constructed with {\sc daophot} a list of 
stars by searching emission 
features on the 5300 \AA\ continuum image. We then used this list of 
stars as input for psf-fitting on the \hei\ and \heii\ images. 
A subtracted image is then built automatically when the photometry 
of all sources is known.
The few remaining faint stars were either removed individually or not
considered for flux determinations. Figs. \ref{br40aha}, \ref{ab7ha} and \ref{ha44} present complete \ha\ images of BAT99-49, AB7 and N44C.
Figs. \ref{br40a} and \ref{n44c} further show close-ups on BAT99-49 and 
N44C in the four nebular filters, while Fig. \ref{ab7} compares the whole
VLT field centered on AB7 in each nebular filter.
Three color images of the high excitation nebulae are available 
on the web at http://www.eso.org/outreach/press-rel/pr-2003/pr-08-03.html 

During the same observing run, we also obtained long-slit spectra of 
these 4 objects with the same instrument used in spectroscopic mode.  
We also made a small survey to study the environment of seven other 
WR stars of the LMC: BAT99-8, 9, 11, 52, 63, 84, and 134.
We used the 600B and 600V grisms to obtain a blue spectrum covering 
the range 3700-5600 \AA\ ($R\sim800$) and a red 
spectrum covering 4500-6850\AA\ ($R\sim1000$), respectively. The 
1.3\arcsec$\times$6.8$'$ slit was tilted, with respect to the N-S 
direction, by 45$^{\circ}$ for the observations of BAT99-2, 11, 
and 134; by 55$^{\circ}$ for BAT99-63; and by 90$^{\circ}$ for AB7. 
In the remaining cases, the slit was oriented in the N-S direction. 
The spatial resolution was 0.9\arcsec-1.4\arcsec\ and the spectral 
resolution, as mesured from the FWHM of the calibration lines, 7\AA. 
The spectra were reduced and calibrated in a standard way using {\sc 
iraf}. For flux calibration, we observed several standard stars from 
Oke (\cite{oke}) and we used the mean atmospheric extinction 
coefficients for CTIO reduced by 15\%. We checked that this 
calibration produces the correct fluxes for the standard stars and 
for a standard Planetary Nebula (Dopita \& Hua
\cite{dop97}). For the 4 high excitation nebulae, 
sky subtraction was done using a small region of the spectra where 
the nebular emission is the lowest. For the WR survey, whenever possible,
we used several regions of lowest emission situated on each side of 
the star and its associated nebula to increase the S/N. Only a few 
residuals remained for the brighter sky lines (e.g. \oi\ 5577\AA). 
The nebular lines were then fitted by gaussians using {\sc splot}.
The Balmer decrement \ha/\hb\ was used to derive the interstellar 
extinction, assuming the theoretical case B decrement (Storey 
\& Hummer \cite{sto}) at either $T$=10~kK, 12.5~kK or 15~kK,
whatever was the closest to the temperature derived using the 
\oiii\ lines (see below). Some spectra were contaminated by faint stars, and
in such cases, we correct the measured emission line strengths by a 2 
\AA\ equivalent width for the Balmer absorptions (McCall et al. 
\cite{mcc}) before estimating the reddening. 
To deredden the line ratios, we used the extinction law from Cardelli 
et al. (\cite{car}) with $R_{{\rm V}}=3.1$ for the LMC (Fitzpatrick \cite{fit}) 
and $R_{{\rm V}}=2.7$ for the SMC (Bouchet et al. \cite{bou}). 
We used these line ratios in conjunction with the {\sc nebular} 
package under {\sc iraf} (Shaw \& Dufour \cite{sha}) to derive 
temperatures, densities and abundances, the latter being 
calculated assuming a constant $n_{{\rm e}}=100\,{\rm cm}^{-3}$ and 
$T_{{\rm e}}$(\oiii). If several lines of the same ion exist, the 
abundance presented in this paper is an average of the abundances 
derived for each line.

\section{The high excitation nebulae}

\subsection{LMC BAT99-49}

Nebular \heii\ emission was discovered near BAT99-49 
(Brey 40a, Sk-7134) by 
Niemela et al. (\cite{nie91}). This emission apparently covered 
70\arcsec\ on their slit, and corresponds to `a partial ring 
seen in \ha\ and \oiii'. In a study of kinematics of WR nebulae, 
Chu et al. (\cite{chu99}) found evidence of interaction between the stellar 
wind of BAT99-49 and the ambient ISM but did not detect a well-defined 
expanding shell. 

   \begin{figure} 
   \centering
   \caption{FORS continuum subtracted \ha, \oiii, \heii, 
and \hei\ images of LMC BAT99-49 and its close surroundings. The images 
are 180\arcsec$\times$180\arcsec. A white cross indicates the 
WR star's position. The crowdiness of this field renders star-subtraction
uttermost difficult. North is up and East to the left.}
              \label{br40a}
    \end{figure}

BAT99-49 has been recently re-classified as WN4:b+O8V by Foellmi et al. 
(\cite{foelmc}). We present in Fig. \ref{br40aha} an \ha\ image of the whole
field, and in Fig. \ref{br40a} a close-up in the four nebular filters. 
To the south of the star, a small arc of 16\arcsec\ radius is visible 
in \oiii, and to a lesser extent in \ha. 
A very bright \ha\ region, BSDL 1985, is also present some 40\arcsec\ south of the 
star. It was identified as the ring nebula associated with BAT99-49 by
Dopita et al. (\cite{dop}). The southern limit of that region has a lower
\oiii/\ha\ ratio than the direct surroundings of BAT99-49. 
A second \ha\ region encloses the star
to the east, at roughly 36\arcsec, but this region completely disappears
in \oiii. A beautiful ring nebula, BSDL 1935 (Bica et al. \cite{bic}) 
is present to the southwest of the field and possesses a very low
\oiii/\ha\ ratio. Generally, the regions to the west of BAT99-49 present
a higher \oiii/\ha\ ratio than those to the east. As for BAT99-2 (see 
Paper I), the \hei\ image appears well correlated with \ha. 
On the contrary, the \heii\ image shows completely different features. The 
brightest \heii\ regions are the small arc 16\arcsec\ south of the star 
and the bright \ha\ region 40\arcsec\ south of the star. A halo surrounds 
the star and fills the region between the star and the two brighter 
features. Still fainter emission is also present, but even our 2700s 
exposure just barely detects it. Compared to BAT99-2, the \heii\ emission
is much fainter but also more extended.

\begin{table} 
\begin{center}
\caption{Average abundances in the Magellanic Clouds, from Russel \& Dopita (\cite{rus}).
\label{abundmc}} 
\begin{tabular}{l c c c c c} 
\hline\hline
 & He/H& O/H & N/H & N/O & S/H\\ 
\hline
LMC & 0.091& 2.3$\times 10^{-4}$& 1.2$\times 10^{-5}$& 0.050&7.4$\times 10^{-6}$\\
SMC & 0.083& 1.3$\times 10^{-4}$& 3.5$\times 10^{-6}$& 0.026&6.8$\times 10^{-6}$\\
\hline
\end{tabular}
\end{center}
\end{table}

 By comparing the flux in ADU on the images at the position of the 
slit with our calibrated spectrophotometry convolved with the filter 
transmission, we were able to calibrate our \heii\ $\lambda$4686 images. 
We then measured the \heii\ $\lambda$4686 flux\footnote{Due to 
errors in the flux calibration and to possible contamination from
a poor subtraction of some stars, we estimate that, throughout 
this paper, the \heii\ fluxes can be considered to have a maximal 
relative error of 50\%.} in a region 36\arcsec$\times$28\arcsec encompassing 
the WR star and the brightest \heii\ emission regions. We find a flux
of about 1$\times$10$^{35}$ erg cm$^{-2}$ s$^{-1}$ after a reddening 
correction $A_{{\rm V}}$ of 0.4 mag (estimated using the Balmer 
decrement, see Sect. 2 and Table \ref{lineratwr}, and assuming a 
uniform reddening). That corresponds to an ionizing flux of 
$\sim$1$\times$10$^{47}$ photons s$^{-1}$. At the LMC metallicity
($Z=0.2-0.4\,Z_{\odot}$), this is what can be expected for a 90-100~kK WN 
star  (Smith at al. \cite{smi}). This value may be a lower limit, 
since we did not take into account all the faintest emission, but 
its contribution should be rather low. With only 5$\times$10$^{36}$ 
He$^+$ ionizing photons s$^{-1}$ (Smith at al. \cite{smi}), the 
O8V companion of the WR star has a negligible influence on the 
\heii\ nebula.

Spectra were extracted from three regions of interest for detailed 
analysis (Fig. \ref{br40aha}): the first one samples the north of 
the star (11\arcsec\ long without the star); the second 
examines the close 15\arcsec\ south of the star and the third one 
explores the bright \ha\ region further south (38\arcsec\ long). 
None contains the WR star, nor another very bright star.
The extinction-corrected line ratios and the physical properties derived 
for these regions are listed in Table \ref{lineratwr}. 
To get helium 
abundances, we used the emissivities at 12500K computed by Benjamin et al.
(\cite{ben}) for \hei\ and by Storey \& Hummer (\cite{sto}) for
H and \heii. The presence of \heii\ suggests that higher ionization
states than O$^{2+}$ exist as well, and we correct the 
$N({\rm O})/N({\rm H})$ ratio
like we did in Paper I. Compared to the LMC 
abundances (Russell \& Dopita \cite{rus}, see Table \ref{abundmc}), no sign of enrichment in 
helium can be detected for the \heii\ nebula. But while the oxygen 
abundance of the bright \ha\ region appears consistent with the LMC, 
the abundance closer to the star is slightly lower, and the N/O ratio
higher than the LMC value. The close neighbourhood of
the star appears thus enriched by a stellar wind displaying CNO
processed material.

\subsection{Star SMC AB7 and its associated nebula SMC N76}

N76 is a large circular nebula some 130\arcsec\ in radius situated to 
the north of the SMC. Nebular \heii\ emission in N76 was first discovered 
serendipitously by Tuohy \& Dopita (\cite{tuo}), while studying the 
neighbouring SNR 1E0101.2-7218. It was rediscovered later by 
Pakull \& Motch (\cite{pak89b}) who noted that the center of gravity of
the \heii\ emission coincides well with the WR star AB7.
Spectra taken by Niemela et al. (\cite{nie91}) show a \heii\ region of 
size 144\arcsec, that fills a `central hole seen in \hei\ and \ha'. 
Because of the presence of the \heii\ nebula, this star, like BAT99-2, was 
classified as WN1 by Pakull (\cite{pak91b}). More recent studies of the 
star undertaken by Niemela et al. (\cite{nie02}) and by Foellmi et al. 
(\cite{foesmc}) led to a re-classification of the star as WN2+O6I(f) 
and WN4+O6I(f), respectively. 

Fig. \ref{ab7ha} shows an \ha\ image of the whole field. N76 nearly completely
fills the FORS CCD. Its surface brightness is not totally uniform: the 
nebula brightens to the east, most probably indicating a density 
gradient. No `central hole' is readily apparent, but many filamentary 
features cover the whole nebula. N76 also presents pillar-like 
features some 92\arcsec\ west of AB7, a quite rare type of structures,
suggesting that the nebula is situated on the edge of a molecular 
cloud being photoevaporated. These structures  may hide a 
second generation of stars (Walborn \cite{wal}).
We also note the presence, to the southeast of AB7, of a bright 
\ha\ knot of diameter 16\arcsec, N76A or DEM S123, which we will
discuss more below.

The \hei\ emission 
appears well correlated with \ha, whilst the \oiii\ image shows 
several differences: mainly, the SNR is clearly visible to the 
northeast of the field, and N76 is slightly smaller in this line 
(124\arcsec\ radius). Some of the filamentary features, the pillars
 and the bright \ha\ emitter DEM S123 present a lower \oiii/\ha\ 
ratio than its surroundings. On the other hand,
the \heii\ image shows a bright elliptic nebula of size 
128\arcsec$\times$146\arcsec. Like in \ha, the surface brightness is 
not uniform and the nebula appears composed of intrincate filaments
making a reversed `S'. These filaments also appears in \ha, but they
are rather faint. More extended, fainter \heii\ emission is also
detectable, corresponding more or less in size with the \ha\ nebula. 

   \begin{figure} 
   \centering
   \caption{FORS continuum subtracted \ha, \oiii, \heii, 
and \hei\ images of SMC N76. The images 
are 394\arcsec$\times$394\arcsec. A white cross indicates the 
position of SMC AB7 (note that the star is wrongly identified
in Dopita et al. \cite{dop}). The SNR is the feature to the NE
of the field only visible in \oiii. North is up and East to the left.}
              \label{ab7}
    \end{figure}

Garnett et al. (\cite{gar91b}) found an average 
$I$(\heii\ $\lambda$4686)/$I$(\hb) of 0.13 at their slit position, but 
we find that this ratio can actually reach higher values locally (see below). 
Assuming a spherical geometry, the same authors derived a luminosity 
$L$(\hb) of 1.1$\times$10$^{38}$ erg s$^{-1}$ and 
$L$(\heii\ $\lambda$4686)=5.7$\times$10$^{36}$ erg s$^{-1}$ for a 
distance of 78 kpc, or $L$(\hb)=6.2$\times$10$^{37}$ erg s$^{-1}$ 
and $L$(\heii\ $\lambda$4686)=3.2$\times$10$^{36}$ erg s$^{-1}$ for our 
adopted SMC 
distance of 59 kpc. Another estimate was made by Pakull \& Bianchi 
(\cite{pak91a}), apparently using spectrophotometry and the same 
approximations as Garnett et al. (\cite{gar91b}). They found a flux in 
\heii\ $\lambda$4686 of 2.5$\times$10$^{-12}$ erg cm$^{-2}$ s$^{-1}$, or 
$L$(\heii\ $\lambda$4686)=1$\times$10$^{36}$ erg s$^{-1}$ for a distance 
of 59 kpc. Using our images calibrated with our FORS spectrophotometry, 
we find a \hb\ luminosity, excluding DEM S123, of 
3.0$\times$10$^{37}$ erg s$^{-1}$ inside a circular aperture of 
130\arcsec\ in radius and a \heii\ $\lambda$4686 luminosity of 
4.5$\times$10$^{36}$ erg s$^{-1}$ inside an elliptic aperture of 
120\arcsec$\times$152\arcsec, after a reddening correction of 
$A_{{\rm V}}$=0.3 mag and for a distance of 59 kpc. 
Since a fainter halo is present, we estimate that the \heii\ 
$\lambda$4686 flux may still be underestimated by 10\%.
Our \heii\ fluxes are larger than those obtained in the previous 
studies, but we use the actual surface brightness' distribution
of the whole nebula, not an extrapolation of a one-position 
measurement. Our measured \hb\ and \heii\ $\lambda$4686 fluxes 
correspond to ionizing fluxes $Q({\rm H}_0)=6.5\times10^{49}$ 
photons s$^{-1}$ and $Q({\rm He}^+)=5.1\times10^{48}$ photons 
s$^{-1}$. In the models of Smith et al. (\cite{smi}), these values 
indicate a WN star hotter than their hottest model, of temperature 
120~kK (for $Z=0.05-0.2\,Z_{\odot}$, $Q({\rm He}^+)=3-4\times10^{48}$ 
photons s$^{-1}$). We know that 
some O stars within N76 could contribute to the ionization of the H 
nebula (the companion of AB7, an O6If star; AzV 327, an O9.7Iab star;
[MWD2000] h53-91 and 137 (Massey et al. \cite{mas}), 
O8.5 V and III stars , respectively), 
but according to the same models, the O-type stars certainly 
do not account for the He$^+$ ionization: at most, 
these stars would contribute by 8.5$\times$10$^{49}$ H ionizing photons 
s$^{-1}$ and 3$\times$10$^{44}$ He$^+$ ionizing photons s$^{-1}$. 
The \heii\ nebula remains a puzzle: either AB7 is an ultra-hot WN star,
or the models need to be adapted or revised, or another source of 
\heii\ ionization is present. In this respect, we note that AB7 does not
appear as a particularly bright X-ray source (Foellmi et al. 
\cite{foesmc}), casting doubts on an X-ray ionization contribution.
Another possibility to produce a high excitation nebula requires
high velocity shocks (Garnett et al. \cite{gar91b}). However, so far 
no kinematic study of N76 was undertaken, and we thus ignore 
whether any high-velocity motions are present. High dispersion
spectroscopy of N76 would thus be valuable in order to better
understand this peculiar object.

   \begin{figure} 
   \centering
   \caption{Upper part: FORS continuum subtracted \ha\ and \oiii\ 
images of SMC N76A (or DEM S 123). Bottom part: \oiii/\ha\ ratio and 
image taken with the continuum filter centered on 6665 \AA. The images 
are 32\arcsec$\times$32\arcsec. North is up and East to the left.
Contrary to the other figures of this paper, the greyscale is a 
linear scale, not a square root scale.}
              \label{dems123}
    \end{figure}

We select spectra from five regions, avoiding the brightest 
stars, for further detailed investigation (Fig. \ref{ab7ha}): three cover
the central \heii\ emission region (W1, 49\arcsec\ long; E1, 
14\arcsec\ long; and E2, 38\arcsec\ long), while the other two explore 
the outer parts of the nebula where \heii\ is barely visible but 
not measurable (W2, 78\arcsec\ long 
and E3, 36\arcsec\ long). The extinction-corrected line ratios for 
these regions are listed in Table \ref{lineratwr}. In the same way 
as for BAT99-49, we then derived temperatures, densities and 
abundances in these regions. No significant enrichment in 
helium can be detected for the \heii\ nebula but the oxygen 
abundance is at least 35\% lower, and the N/O ratio is
at least 40\% higher than the SMC's average values (Russell \& Dopita 
\cite{rus}, see Table \ref{abundmc}). Like in BAT99-2 and 49, this suggests a small enrichment
by stellar winds, but in the case of N76, the nebula over which this
enrichment is seen is much larger. 

\subsubsection{LMC N76A (or DEM S 123)}

To the southeast of AB7 is situated N76 A (or DEM S 123, see Fig. \ref{dems123}), a bright \ha\ knot of radius $\sim$8\arcsec\ 
similar to compact \hii\ regions such as N11A in the 
LMC (Heydari-Malayeri et al. \cite{hey}). Following Heydari-Malayeri 
et al. (\cite{hey} and references therein), these compact regions are 
created by massive stars just leaving their natal molecular cloud.
The same authors also called this type of objects `High Excitation 
Blobs' (HEBs). In N76A, one very bright star lies
near the center of this nebula, and five others are situated closer 
to its periphery (see Fig. \ref{dems123}). Unfortunately, the spectral 
types of these stars 
are unknown and may be worth further investigation. The brightest 
region of N76A is situated just to the NE of the brightest star, 
whilst a region of suppressed brightness is directly symetrical to it
with respect of the star. A diffuse envelope with few remarkable 
features surrounds the center of this nebula, and a dust lane borders 
it to the east. 

As mentioned above, the \oiii/\ha\ ratio of N76A is much lower than that of 
its direct surroundings. It is actually the lowest to the northern
and eastern edges of the nebula (\oiii/\ha$\sim$0.3), while 
it peaks to the southwest of the brightest star (\oiii/\ha$\sim$0.7). 
In a similar way as above, we estimate the 
\ha\ and \oiii\ fluxes in an aperture of 8\arcsec\ radius and find 
an observed $L$(\ha)$\sim1\times10^{36}$ erg s$^{-1}$ and an observed 
$L$(\oiii\ $\lambda$5007)$\sim6\times10^{35}$ erg s$^{-1}$ for a 
distance of 59 kpc. However, since the reddening of this region is 
not known, we thus cannot derive the actual luminosities. 
But the observed \ha\ luminosity can still be used to estimate the 
minimum requirement on the spectral type in the case of a unique 
ionizing star: it corresponds to an ionizing flux of 
$\sim$7$\times$10$^{47}$ ionizing photons s$^{-1}$, which
can be explained by an O9V type star. Further investigations 
should enable to find the reddening of the nebula and the exact nature 
of its embedded stars, but also to check the predicted correlation 
between the \oiii/\hb\ ratio and the \hb\ flux for the HEBs
(Heydari-Malayeri et al. \cite{hey}).

   \begin{figure*} 
   \centering
   \caption{FORS \ha\ image of LMC N44. 
The different regions used for spectral analysis 
are marked by a solid line. Features discussed in the text 
are labelled.}
              \label{ha44}
    \end{figure*}

\subsection{LMC N44C}

N44C is a bright elongated \ha\ nebula some 53\arcsec$\times$65\arcsec\ 
situated to the southwest of a superbubble blown by the OB
association LH47 (see Fig. \ref{ha44}). 
Nebular \heii\ $\lambda$4686 emission was detected in N44C by 
Stasi\'nska et al. (\cite{sta}) who also provided the first spectra 
of this nebula. Examining a high dispersion \ha\ spectrum of N44C, Goudis 
\& Meaburn (\cite{gou}) claimed to have discovered gas moving at $-$120\kms\
in this nebula, but it was found later that this component was actually 
`only' the \heii\ $\lambda$6560 line (Garnett et al. \cite{gar00}). 
Two bright stars in the center of N44C have been detected by Stasi\'nska 
et al. (\cite{sta}) and labelled Star \#2 for the northern one and Star 
\#1 for the southern one. Star \#2 is a rather hot O-type star while
Star \#1 is a cool G-K foreground star (Stasi\'nska et al. \cite{sta}, 
Garnett et al. \cite{gar00}). 

In the absence of high velocity motions or any very hot star, 
it was suggested that the He$^+$ ionization results from an intense 
X-ray source inside the nebula. Such a process is actually seen at work 
around another
object of the LMC, LMC X-1, and in this case, the \heii\ nebula 
is complex and very extended. If the X-ray source is transient, the
high excitation may last for some time after the X-ray source has
switched off. For N44C, the \heii\ nebula has been 
proposed to be a fossil X-ray nebula corresponding to the 
once recorded X-ray transient LMC X-5 (Pakull \& Motch \cite{pak89a}). 
An extensive study of N44C was recently presented by Garnett et al.
(\cite{gar00}) who evaluated that \heii\ is currently recombining with a 
characteristic timescale of 20 yr. Since their spectrum was taken in 1991
and Stasi\'nska's in 1985, it was obviously interesting to monitor the 
evolution of the \heii\ emission ten to fifteen years later. 

   \begin{figure} 
   \centering
   \caption{FORS continuum subtracted \ha, \oiii, \heii, 
and \hei\ images of LMC N44C and its close surroundings. The images 
are 200\arcsec$\times$200\arcsec. The upper white cross indicates 
the Star \# 2 of Stasi\'nska et al. (\cite{sta}), and the lower one
shows the position of Star \# 1. North is up and East to the left.}
              \label{n44c}
    \end{figure}

\begin{figure} 
   \centering
   \includegraphics[width=9cm]{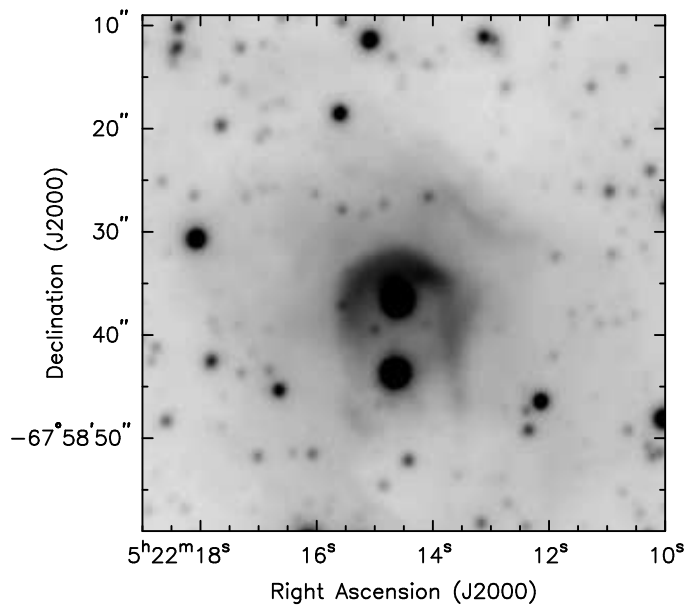}
   \caption{Close-up on LMC N44C in the \heii\ $\lambda$4686 line
(the image is {\em not} continuum subtracted). Stars \# 1 and 2 are 
clearly visible, as is a bright arc-like \heii\ emission region 
enveloping the Star \# 2. Fainter, more extended emission also exists, 
but is more apparent on Fig. \ref{n44c}.}
              \label{n44czoom}
    \end{figure}

Images of N44C in the four nebular filters are presented in Fig. \ref{n44c}.
The \ha\ and \oiii\ images have been described in length by Garnett et
al. (\cite{gar00}) and we will not repeat the discussion here.
The \hei\ image correlates perfectly with \ha, except for a slightly 
\hei\ depleted region where the \heii\ emission is the brightest.
The \heii\ image presents a very bright arc north to Star \#2 and 
a fainter, more extended emission. A close-up on the arc, whose radius 
is $\sim$6.5\arcsec, is presented in Fig. \ref{n44czoom}. It seems to
closely envelope the two stars. This arc does not correspond to any 
feature in the other nebular lines. On the contrary, the halo is as 
extended as the \ha\ nebula, and well correlated with it. However, 
an important part of this halo is actually made of \ariv\ emission:
in the spectra, the observed \ariv\ $\lambda$4711 + 
\hei\ $\lambda$4713 blend can reach 
four times the intensity of the \heii\ $\lambda$4686 line.
In the rest of the field, emission in the \heii\ filter is spotted in N44B, 
N44H and in the superbubble. However, these features are most probably 
not due to a nebular \heii\ emission. In fact, our spectrum reveals a 
faint \feiii\ $\lambda$4658 line in the superbubble and
Dufour (\cite{duf}) detected \feiii\ $\lambda$4658 in 
N44B, although Stasi\'nska et al. (\cite{sta}) think that
Dufour actually observed N44C, not B, and mistook \heii\ 
$\lambda$4686 for the \feiii\ line. Unfortunately, we do not have any 
information in the case of N44H's emission, since our spectra do not 
cover this region.

We choose six regions to further examine the physical properties 
of the nebula and its surroundings (see Fig. \ref{ha44}): the 
\heii\ arc situated directly north of Star \#2 (N1, 4.6\arcsec\ 
long), the northern part of N44C (N2, 15\arcsec\ long), the region 
directly south of Star \#1 (S1, 11\arcsec\ long), the southern 
part of N44C that seems composed of interlaced arcs (S2, 
25\arcsec\ long), a filament just north of N44C which possesses 
an ionization closer to N44C than to the superbubble (N Fil., 
5.2\arcsec\ long), and finally the superbubble itself (Sup., 
13\arcsec\ long). The dereddened line ratios of these six regions, 
together with the derived physical properties, are presented in 
Table \ref{lineratn44c}. The helium abundance is close to normal in all 
cases, except for the superbubble where it is rather low. The 
oxygen abundance is at most slightly smaller than normal, but the 
N/O ratio always reflects the LMC average (see Table \ref{abundmc}). 
No significant chemical enrichment is thus detected for this nebula. 

We estimate the \ha\ and \heii\ fluxes by calibrating 
the corresponding images with our spectrophotometry. For the
whole \ha\ nebula (53\arcsec$\times$65\arcsec), we derived
a \hb\ luminosity of 9$\times$10$^{36}$ erg s$^{-1}$ after
a reddening correction $A_{{\rm V}}=0.55\,{\rm mag}$ and for a distance of 50kpc.
 Note that the \hb\ luminosity given in Garnett et al. 
(\cite{gar00}) is an order of magnitude larger than this value. 
However, this is most probably a typographical error: Garnett et al. 
(\cite{gar00}) quote a reddening-corrected \hb\ 
flux of 3$\times$10$^{-10}$ erg cm$^{-2}$ s$^{-1}$ for the sole 
N44C, while Caplan \& Deharveng (\cite{cap}) give an observed 
flux of 1.1$\times$10$^{-10}$ erg cm$^{-2}$ s$^{-1}$ for N44 B + C 
(N44B is more extended and as bright as N44C !). The analysis 
of the HST images used by Garnett et al. (\cite{gar00}) 
and several other calibrated \ha\ images of this nebula 
confirm our flux estimate\footnote{Most of the 
differences in $L$(4686)/$L$($\beta$) between Garnett et al. 
(\cite{gar91b}) and Garnett et al. (\cite{gar00}) - which were discussed
in length in the latter - can be explained 
by this error. Further discrepancies can be explained by the fact 
that neither the \heii\ nor the \ha\ nebula presents a spherical 
shape; that Stasi\'nska et al. (\cite{sta}) measured only the 
northern edge of the \heii\ nebula and that there may be an error in
the Stasi\'nska et al. (\cite{sta}) paper (Pakull \& Motch \cite{pak89a}). 
We also note that our extinction $A_{{\rm V}}$=0.55 mag, corresponding to $c_{\beta}=0.26$ and a color excess $E(B-V)=0.18\,{\rm mag}$, is close to 
the average value found by Oey \& Massey (\cite{oey}). It is thus possible
that the extinction $A_{{\rm V}}=0.25\,{\rm mag}$ of Garnett et al. 
(\cite{gar00}) actually refers to $c_{\beta}$. However, that fact alone
can not explain the large discrepancy observed between the \heii\ fluxes.}. 
Our \hb\ luminosity corresponds to an ionizing flux of 1.9$\times$10$^{49}$ 
photons s$^{-1}$. Due to the pollution by the nebular lines and 
to possible abundance anomalies, the spectral type of Star \#2 
is uncertain. Pakull \& Motch (\cite{pak89a}) proposed an O4-6 type, 
while Garnett et al. (\cite{gar00}) prefered ON7 III-V. The observed
$Q({\rm H}_0)$ is a bit too large for a O7V star and would better fit
to a slightly earlier main sequence O-type star, but it is compatible 
with an O7III type. However, the magnitudes given in Stasi\'nska et al.
(\cite{sta}) and Oey \& Massey (\cite{oey}) seem to favor a main sequence 
classification.

When trying to estimate the \heii\ flux, we discovered an apparent 
continuum emission to the north of Star \#2 in our spectra. The \heii\ 
image thus actually contains \heii+\ariv+cont., and the pollution due to 
these additional components is larger where the \heii\ nebula is fainter.
That is the reason why we choose to measure the \heii\ flux only 
in the brightest parts of the nebula, i.e. a region of radius 7\arcsec\
comprising the bright arc. Therefore we may miss part of the \heii\ flux, 
but part of this error is compensated by the contamination of \ariv\ + cont.
We find $L$(\heii\ $\lambda$4686)$=4\times10^{35}$ erg s$^{-1}$ and 
$Q({\rm He}^+)=4\times10^{47}$ photons s$^{-1}$, after the reddening 
correction and for a distance of 50kpc. In the past, a few estimates 
of the \heii\ flux have been made: using the results of Stasi\'nska et 
al. (\cite{sta}), Garnett et al. (\cite{gar91b}) found $L$(\heii\ $\lambda$4686)$=6.3\times10^{35}$ erg 
s$^{-1}$ for a distance of 57 kpc (or $4.7\times10^{35}$ erg s$^{-1}$ 
at a distance of 50kpc) but their estimate was made considering a 
spherical geometry, which is obviously not the case. This approximation
could lead to a large overestimate of the flux, but since Stasi\'nska 
et al. (\cite{sta}) did not measure the $I$(\heii\ $\lambda$4686)/$I$(\hb) 
in the brightest part of the \heii\ nebula, their result is probably 
not too far off. Using again Stasi\'nska et al. (\cite{sta}) measurements, 
Pakull \& Motch (\cite{pak89a}) quote a lower value $L$(\heii\ 
$\lambda$4686) of $3\times10^{35}$ 
erg s$^{-1}$, after `taking into account an apparent error in 
Stasi\'nska et al. (\cite{sta}) paper'. Garnett et al. (\cite{gar00}) 
found $L$(\heii\ $\lambda$4686)$=2.2\times10^{35}$ erg s$^{-1}$, which is 
lower than our value, but the region considered by these authors
is not specified clearly, and their \heii\ images, taken with the 
CTIO 0.9m telescope, probably did not 
reach a high signal/noise (the authors actually did not show the images). 
However, since we took our spectra with the same slit orientation as in 
Garnett et al. (\cite{gar00}), we can compare rather the $I$(\heii\ $\lambda$4686)/$I$($\beta$) ratio.
Garnett et al. (\cite{gar00}) quoted a value of 0.064$\pm$0.003, and we 
find 0.061$\pm$0.004 when considering like them the whole nebula, 
including Stars \# 1 \& 2. Furthermore, in a region similar and as 
close as possible to the position observed by Stasi\'nska et al. 
(\cite{sta}), we get similar results as these authors as well. We thus 
notice hardly any long-term change in the \heii\ emission. 

Apparently, the \heii\ emission of N44C is still a puzzle. The ionizing 
star(s) in the vicinity cannot provide enough hard photons and there 
is no high-velocity gas that could trace a shock capable of ionizing 
He$^+$. The last hypothesis - a fossil X-ray nebula (Pakull \& Motch 
\cite{pak89a}) - implies a recombination of \heii, while we see no clear evidence for a change in the line intensity over the last 15 years. 
Recombination of \neiv\ and \nev\ is also expected, but unfortunately, 
there are no lines of these ions in the wavelength ranges covered by
our spectra. Pakull \& Motch (\cite{pak89a}) estimated
recombination timescales of 20 and 100 yr for \nev\ and \heii, 
respectively. Garnett et al. (\cite{gar00}) favored smaller
values, of 5 and 20 yr\footnote{However, we note that Garnett et al. 
(\cite{gar00}) were puzzled by the strength of the observed \neiv\ 
lines, incompatible with their short timescales.}.
Since the proposed associated X-ray source, LMC X-5,  was 
observed in 1974-76 by OSO7 and Ariel V, we should detect a 
significant reduction of the \heii\ lines, even with a timescale 
as long as 100 yr. 
Apart from a fourth and still unknown ionizing mechanism, one 
possibility to explain the observations could be that the recombination 
timescale is much longer, thus that the density in N44C is very low. 
Another could be that the mysterious X-ray source responsible of the 
ionization had recently turned on again, reionizing the nebula, but we 
note that Einstein data, a ROSAT pointing, and a recent $Chandra$ 
observation of N44 do not
reveal any source in the vicinity of N44C (Y.-H. Chu, private communication).
It would be worth to follow the evolution of the \heii, \neiv\ and 
\nev\ and X-ray emissions in the future, to determine the temporal behaviour of the nebula. The shape and the location of the \heii\ emission strongly
suggest that the phenomenon is associated in some way with Star \#2.
Monitoring of this object should allow to detect signs of orbital motion, 
and thus the presence of the hypothetical compact companion needed for 
the fossil X-ray nebula scenario. However, we note 
that the 90\% confidence error on the position of LMC X-5 is 
$\sim$0.3$^{\circ}$ while N44C lies 
at 0.5$^{\circ}$ from the X-ray source.
We further note the absence of any evidence (in the X-ray or radio 
domains, by means of high velocity motions or enhanced \sii/\ha\ 
ratio) of a supernova explosion in the vicinity of Star \#2 that 
could have given birth to a compact object.

\begin{table*} 
\begin{center}
\caption{Same as Table \ref{lineratwr} for LMC N44C. \label{lineratn44c}} 
\begin{tabular}{l c c c c c c} 
\hline\hline
 & N1& N2& S1& S2& N Fil. & Sup.\\ 
\hline
\feii\ 3712& & & & & & 9.1 (1.)\\
\oii\ 3727& 91 (9) & 220 (23) & 199 (21) &353 (37) & 547 (57) & 486 (51)\\
H12& 4. (0.4) & 3.7 (0.4) & 4.7 (0.5)&4.9 (0.5) & 4.6 (0.5) & 5.6 (0.6)\\
H11& 4.8 (0.5) & 4.6 (0.5) & 6.2 (0.6)& 5.9 (0.6) & 6.3 (0.7) & 6.4 (0.7)\\
H10& 5.9 (0.6) & 5.8 (0.6) & 7.4 (0.8)& 7. (0.7) & 6.8 (0.7) & 6.1 (0.6)\\
H9& 7.5 (0.8) & 7.8 (0.8) & 8.8 (0.9) &9. (0.9) & 8.3 (0.8) & 8.1 (0.8)\\
\neiii\ 3868& 63 (6) & 73 (7) & 59 (6) &56 (6) & 65 (6) & 11 (1)\\
H8 + \hei& 19 (2) & 21 (2) & 20 (2)& 20 (2) & 19 (2) & 20 (2)\\
\neiii\ + \he & 35 (3) & 39 (4) & 34 (3) &35 (3) & 35 (3) & 22 (2)\\
\hei\ 4026 & 2.1 (0.2) & 2.2 (0.2) & 1.7 (0.2)& 1.7 (0.2) & 1.8 (0.2) & 1.3:\\
\sii\ 4068 & 1.3 (0.1) & 2.7 (0.3) & 2.4:& 4. (0.4) & 6.7 (0.6)& 4.3:\\
\sii\ 4076 & 0.4 (0.2) & 0.5 (0.2) & & 0.6 (0.1) & 1. (0.2)&\\
\hd & 29 (3) & 29 (3) & 30 (3) & 30 (3) & 29 (3) & 29 (3)\\
\hg & 50 (4) & 50 (4) & 50 (4) & 50 (4) & 50 (4) & 50 (4)\\
\oiii\ 4363& 7.4 (0.6) & 7.3 (0.6) & 6.7 (0.5) &4.9 (0.4) & 5.5 (0.4) & 0.8 (0.2)\\
\hei\ 4389 & 0.4 (0.2) & 0.5 (0.3) & & & &\\
\hei\ 4471& 3.6 (0.3) & 4.4 (0.3) & 3.8 (0.3)& 3.7 (0.3) & 4.1 (0.3) & 3.6 (0.3)\\
\heii\ 4542 & 0.4 (0.2) & & & & & \\
\heii\ 4686& 14 (0.1) & 0.45 (0.03) & 4.3 (0.3)& 1.7 (0.1) & &\\
\ariv\ 4711$^a$& 3.3 (0.3) & 1.2 (0.1) & 1.4 (0.1)& & &\\
\ariv\ 4740& 2.5 (0.2) & 1. (0.2) & 1.2 (0.1)&  & &\\
\hb & 100. & 100. & 100.& 100.& 100. & 100.\\
\hei\ 4922 & 0.9 (0.2) & 1.1 (0.1) & 1.1 (0.1) &1. (0.2) & 1.1 (0.1) & 1.6:\\
\oiii\ 4959& 242 (17) & 250 (18) & 217 (15) & 178 (13) & 161 (11) & 36 (3)\\
\feiii\ 4986 &  & & & & & 2.2 (0.2)\\
\oiii\ 5007& 719 (51) & 747 (53) & 649 (46) & 534 (38) & 478 (34) & 108 (8)\\
\no\ 5198+5200 & 0.2 (0.1) & 0.6 (0.1) &  & 1. (0.2) & 2. (0.2)& 1.:\\
\heii\ 5412& 1.1 (0.1) & & 0.7 (0.1) &  & &\\
\cliii\ 5517 & 0.4 (0.2) & 0.5 (0.1) & 0.6 (0.1) & 0.6: & & \\
\cliii\ 5538 & 0.3 (0.1) & 0.4 (0.2) & 0.5 (0.1) & 0.5: & & \\
\nii\ 5755 & 0.13 (0.07) & 0.3 (0.2) & & 0.5 (0.1) & 0.7 (0.1)& \\
\hei\ 5876 & 10 (1) & 12 (1) & 10 (1) & 10 (1) & 9.8 (0.8) & 9.3 (0.1)\\ 
\feiii\ 6096 & & & & & & 1.5 (0.1)\\
\oi\ 6300  & 1.9 (0.2) & 6.8 (0.7) & 3.9 (0.4) & 8.4 (0.8) & 14 (1) & 4.1 (0.4)\\
\siii\ 6312& 1.4 (0.1) & 1.9 (0.2) & 1.6 (0.2) &1.9 (0.2) & 2. (0.2) & 1. (0.1)\\
\oi\ 6363& 0.8 (0.2) & 2.4 (0.2) & 1.4 (0.1) & 2.9 (0.3) & 4.8 (0.5) & 1.8 (0.2)\\
\nii\ 6548 & 2. (0.2) & 5.2 (0.5) & 5. (0.5) &8.7 (0.9) & 14 (1) & 13 (1) \\
\ha$^a$& 282 (28) & 282 (28) & 282 (28) &282 (28) & 282 (28) & 286 (29)\\
\nii\ 6583& 6.5 (0.7) & 16 (2) & 16 (2)& 27 (3) & 42 (4) & 38 (4)\\
\hei\ 6678&  2.9 (0.3) & 3.3 (0.3) & 2.9 (0.3) &2.7 (0.3) & 2.8 (0.3) & 2.6 (0.3)\\
\sii\ 6716& 8.9 (0.9) & 22 (2) & 24 (3)& 42 (4) & 58 (6) & 51 (5)\\
\sii\ 6731& 7. (0.7) & 17 (2) & 17 (2)& 30 (3) & 42 (4) & 36 (4)\\
$F$(\hb) (10$^{-14}$ erg cm$^{-2}$ s$^{-1}$)& 18& 23& 8.4& 13& 1.7&2.4\\
\hline
$A_{{\rm V}}$(mag)& 0.58& 0.60& 0.52& 0.53& 0.48&0.43\\
$T_{{\rm e}}$\oiii\ (kK) &  11.7$\pm$0.4 & 11.5$\pm$0.4 & 11.7$\pm$0.4& 11.3$\pm$0.4& 12.2$\pm$0.4& 10.5$\pm$0.9 \\
$T_{{\rm e}}$\sii\ (kK) &  11.8$\pm$2. & 9.9$\pm$1. & & 8.3$\pm$0.4& 9.3$\pm$0.8& \\
$T_{{\rm e}}$\nii\ (kK) &  12$\pm$5 & 12$\pm$4 & & 11.4$\pm$1.4& 10.7$\pm$1.1&  \\
$n_{{\rm e}}$\sii\ (cm$^{-3}$) & $<$420& $<$410& $<$260& $<$250& $<$270&$<$240\\
$n_{{\rm e}}$\ariv\ (cm$^{-3}$) & $<$2400& $<$5400& 600-3600& & &\\
\hline
\end{tabular}
\end{center}
\end{table*}

\setcounter{table}{3} 

\begin{table*} 
\begin{center}
\caption{Continued.} 
\begin{tabular}{l c c c c c c} 
\hline\hline
(He$^+$/H$^+\times 10^{2}$)$^b$ & 7.7$\pm$0.5 & 9.$\pm$0.6 & 7.9$\pm$0.3& 7.6$\pm$0.4& 7.9$\pm$0.3&6.9$\pm$0.3 \\
He$^{2+}$/H$^+\times 10^{2}$ & 1.2$\pm$0.2 & 0.038$\pm$0.003 & 0.55$\pm$0.08& 0.14$\pm$0.01& & \\
$\rightarrow$ He/H$\times 10^{2}$ & 8.9$\pm$0.5 & 9.1$\pm$0.6 & 8.4$\pm$0.3& 7.7$\pm$0.4& 7.9$\pm$0.3&6.9$\pm$0.3 \\
O$^{0+}$/H$^+\times 10^{6}$ &2.4$\pm$0.3 & 8.3$\pm$0.6 & 4.5$\pm$0.2& 11$\pm$1& 14$\pm$1&7.9$\pm$0.5 \\
O$^+$/H$^+\times 10^{5}$ & 1.8$\pm$0.2 & 4.7$\pm$0.5 & 4.$\pm$0.4& 8.$\pm$0.8& 9.3$\pm$1.&15$\pm$2 \\
O$^{2+}$/H$^+\times 10^{4}$ & 1.5$\pm$0.1 & 1.7$\pm$0.1 & 1.4$\pm$0.1& 1.2$\pm$0.1& 0.89$\pm$0.04 & 0.32$\pm$0.03\\
$\rightarrow$ O/H$\times 10^{4}$ & 2.$\pm$0.1 & 2.2$\pm$0.1 & 1.9$\pm$0.1& 2.2$\pm$0.1& 2.0$\pm$0.1&1.9$\pm$0.2 \\
N$^{0+}$/H$^+\times 10^{7}$ & 0.84$\pm$0.04 & 2.3$\pm$0.5 & &3.8$\pm$0.8& 5.9$\pm$0.5& \\
N$^+$/H$^+\times 10^{6}$ & 0.84$\pm$0.16 & 2.2$\pm$0.4 & 2.$\pm$0.1& 3.9$\pm$0.3& 4.6$\pm$0.4&6.5$\pm$0.5 \\
($\rightarrow$ N/O$\times 10^{2}$)$^a$ & 4.7$\pm$1. & 4.8$\pm$0.9 & 5.2$\pm$0.6& 5.3$\pm$0.7& 4.9$\pm$0.6&4.4$\pm$0.6 \\
(S$^+$/H$^+\times 10^{7}$)$^c$ & 2.6$\pm$0.2 & 5.9$\pm$0.4 & 6.7$\pm$0.5& 10$\pm$1& 13$\pm$1&16$\pm$1 \\
S$^{2+}$/H$^+\times 10^{6}$ & 1.8$\pm$0.2 & 2.6$\pm$0.3 & 2.1$\pm$0.2& 2.8$\pm$0.3& 2.2$\pm$0.2&2$\pm$0.2 \\
Ar$^{3+}$/H$^+\times 10^{7}$ & 3.4$\pm$0.2& 1.3$\pm$0.2 & 1.3$\pm$0.1 & & & \\
Ne$^{2+}$/H$^+\times 10^{5}$ & 3.8$\pm$0.4 & 4.7$\pm$0.5 & 3.6$\pm$0.4& 3.9$\pm$0.4& 3.4$\pm$0.3&0.96$\pm$0.10 \\
Cl$^{2+}$/H$^+\times 10^{8}$ & 2.4$\pm$0.9 & 3.6$\pm$1. & 4.5$\pm$0.7& & & \\
\hline
\end{tabular}
\end{center}
$^a$ Same remarks as in Table \ref{lineratwr}.\\
$^b$ To derive this abundance, the weigth of the fainter \hei\ $\lambda$ 4026, 4389 and 4922 lines was reduced to 0.5. \\
$^c$ To derive this abundance, the weigth of the \sii\ $\lambda$ 4076 line was reduced to 0.5. 
\end{table*}

\section{WR survey}

We have undertaken a small survey to search for \heii\ emission
around a few WR stars of the LMC. Similar surveys 
 had previously been undertaken by Niemela et al. (\cite{nie91})
and Pakull (\cite{pak91b}), mainly concentrating on WN stars. 
We chose seven other WR stars:  BAT99-8, 9, 11, 52, 63, 84
and  134. None of them shows any sign of nebular \heii\ emission, despite 
the fact that Melnick \& Heydari-Malayeri (\cite{mel}) claimed to have 
found possible extended \heii\ emission around BAT99-9. However, BAT99-8, 
11, 63 and 134 are known to present ring nebulae around them, and we study
here their physical properties, which can help to determine the evolutionary 
status of the WR bubbles. Chu et al. (\cite{chu99}) have suggested that
the large size and surface brightness variations of the nebulae around
BAT99-11, 63 and 134 support the existence of some interactions with 
the ISM but also that their large expansion velocities and the rather 
regular expansion pattern show that they are not yet dominated by the ISM.

\subsection{BAT99-8}
BAT99-8 (Brey 8, Sk-6942, HD32257), 
a WC4 star (Bartzakos et al. \cite{bar}), is surrounded
by a ring nebula of size 150\arcsec$\times$230\arcsec\ (Dopita et al. 
\cite{dop}). Two faint, arcuate filaments are visible to the north 
and southeast of the star. They are more easily distinguished 
from the background nebula in the \oiii\ lines. Unfortunately, no 
kinematic study of this candidate wind-blown bubble is available, 
and its exact nature and expansion status are still unknown. 
We have analysed the spectra of each arc (15\arcsec\ long for the 
northern one and 19\arcsec\ for the southern one), and we present 
in Table \ref{lineratwr2} their dereddened line ratios.\\

No nebular \heii\ emission was detected in this nebula, but the 
spectra still allow us to derive an upper limit on the strength of 
the \heii\ emission by estimating the 3$\sigma$ level at the position 
of \heii\ $\lambda$4686. This leads to $F$(\heii\ $\lambda$4686)/$F$(\hb)$<$0.007 in the most restrictive case, i.e.
the one that gives the lowest upper limit. Assuming
a geometry composed of two partial rings of uniform brightness
(a quarter of a ring of radius 65\arcsec\ and width 7\arcsec, plus 
one eighth of a ring of radius 111\arcsec\ and width 7\arcsec, see 
the morphology of the \ha\ nebula on Fig. 1c of Dopita et 
al. \cite{dop}), a distance of 50 kpc and a reddening $A_{{\rm V}}$ 
of 0.4 mag, we then find an upper limit on the ionization power of 
$\log[Q({\rm He}^+)]<45.3$.\\

The faintness of the \oiii\ lines renders the determination of the 
temperature of the nebula rather difficult: in our sample, 
such weakness of the \oiii\ $\lambda\lambda$ 4959, 5007 lines  
compared to \hb\ is only seen for the N44 superbubble. That may suggest 
a temperature of 10~kK for this nebula. Such a low temperature 
is further supported by the rather strong \oii, \sii\ and \nii\ 
lines. For this temperature, we found only an oxygen abundance slightly 
higher than the LMC average. Higher temperatures of 12.5~kK or 15~kK 
could lead to significant departures from average, with a  
low oxygen abundance and a quite high N/O ratio, but such a 
temperature appears less probable.

\subsection{BAT99-11}
The nebula surrounding BAT99-11 (Brey 10, Sk-6815, HD32402), 
a WC4 star (Bartzakos et al. 
\cite{bar}), was discovered quite a long time ago by Chu 
(\cite{chu81}). It has an elliptical shape 
of size 180\arcsec$\times$112\arcsec\ and its surface brightness 
is far from uniform, with the southwestern region being 
the brightest (Dopita et al. \cite{dop}). A kinematic study 
reveals that this nebula corresponds to a shell expanding 
into a quiescent \hii\ region with a velocity of $\sim$42\kms
(Chu et al. \cite{chu99}). We have examined the spectra of 
both sides of the nebula (29\arcsec\ for the SW arc and 
48\arcsec\ for the NE one), and present their line ratios 
in Table \ref{lineratwr2}. 
As above, we estimated upper limits on the strength of the \heii\ 
emission of this nebula and find $F$(\heii\ $\lambda$4686)/$F$(\hb)$<$0.002 
in the most restrictive case. Assuming an elliptical ring geometry 
(111\arcsec$\times$183\arcsec\ for the size of the external ellipse,
and 85\arcsec$\times$144\arcsec\ for the internal one)
and a reddening $A_{{\rm V}}$ of 0.58 mag, we then find $\log[Q({\rm He}^+)]<46.2$.

The large temperature derived for the fainter northeastern 
region is probably due to the rather low S/N in the \oiii\ 
$\lambda$ 4363 line, and is thus to consider with caution. The 
most reliable abundances, derived for the brightest side of the 
nebula, only show O/H and N/O values slightly lower than the LMC 
values. Similar results are found for
the NE region when using the temperature of the SW one, whilst
a normal helium abundance and N/O ratio, but a very low oxygen 
abundance are found for the higher temperature derived from the
\oiii\ lines. Like for BAT99-2, we may be observing a circumstellar 
bubble merging with the interstellar one.

\subsection{BAT99-63}
A fine filamentary ring nebula, some 65\arcsec$\times$88\arcsec\ 
in size, surrounds the star BAT99-63 (Brey 52, classified WN4ha: in Foellmi 
et al. \cite{foelmc}). The southwestern part of the ring is 
the faintest, while the northeastern region is much brighter.
The analysis of the kinematics of this nebula suggests a 
blister structure, with an expansion velocity of $\sim$50\kms\ in the 
low density ISM (Chu et al. \cite{chu99}). 
We studied the spectrum of the brightest region, situated
to the NE of the star (62\arcsec\ long, see Table \ref{lineratwr2}). 
Using the same method as for BAT99-8 and 11, we find an upper limit
on the \heii\ emission of $F$(\heii\ $\lambda$4686)/$F$(\hb)$<$0.006
in the most restrictive case. Assuming an elliptical ring geometry 
(80\arcsec$\times$70\arcsec\ and 58\arcsec$\times$45\arcsec\ for the 
sizes of the external and internal ellipses, respectively)
and a reddening $A_{{\rm V}}$ of 0.4 mag, we then find $\log[Q({\rm He}^+)]<45.4$.
Even if the oxygen abundance might be slightly larger than the LMC 
average, no significant chemical enrichment can be detected in this nebula,
favoring the hypothesis of an interstellar bubble already dominated by 
the ISM.

\subsection{BAT99-134}
A beautiful ring nebula of size 87\arcsec$\times$107\arcsec\ 
(Dopita et al. \cite{dop}) completely surrounds BAT99-134 (Brey 100, Sk-67268, HD270149), 
a WN4b star (Foellmi et al. \cite{foelmc}). Its southeastern side
is brighter, and the WR star appears largely decentred. Using high 
dispersion spectra, Chu (\cite{chu83}) detected an 
expansion velocity of 45-50\kms\ in this nebula. The same 
author proposed that it is actually a small shell expanding on 
the surface of a molecular cloud. She further suggested that the 
motion of the WR star, probably originally formed on the cloud's 
surface, had caused the offset. We decided to analyse four 
regions in this nebula: the first and second sample the NE and SW 
part of the ring (18\arcsec\ and 15\arcsec\ long, respectively), the 
third covers the region situated between the star and the SW arc 
(47\arcsec\ long), and the last one studies the \hii\ region beyond 
the SW arc (46\arcsec\ long), that belongs to a larger shell 
unrelated to the WR wind-blown bubble. Compared to the LMC average
values, the regions of the nebula closer to the star and the NE arc
appear slightly enriched in helium and depleted in oxygen. 
The N/O ratio is also generally larger. These abundance differences
fit well the Chu (\cite{chu83}) model: the southwesternmost part
of the wind-blown bubble and the outer \hii\ region actually belong 
to the molecular cloud's edge, and should thus present normal 
abundances. 

Using the same method as above, we find an upper limit
on the \heii\ emission of $F$(\heii\ $\lambda$4686)/$F$(\hb)$<$0.007
in the most restrictive case. Assuming an elliptical ring geometry 
(80\arcsec$\times$100\arcsec\ and 66\arcsec$\times$79\arcsec\ for the 
sizes of the external and internal ellipses, respectively)
and a reddening $A_{{\rm V}}$ of 0.3 mag, we then find $\log[Q({\rm He}^+)]<45.5$.

\subsection{WR stars and nebular \heii\ emission : discussion}

Up to now, only 3 WR stars of the MCs are known to display nebular \heii\ 
emission: BAT99-2, 49 and AB7, while 32 others (BAT99-5, 6, 8, 9, 11, 12, 
15, 19, 29, 31, 32, 36, 38, 43, 52, 53, 56, 59, 60, 62, 63, 64, 81, 82, 
84, 92, 94, 123, 124, 126, 132 and 134, Pakull \cite{pak91b}, Niemela
\cite{nie91}, and this paper) have none. For LMC WC4 stars, including
our studied cases of BAT99-8, 9, 11 and 52, Gr\"afener et al. (\cite{gra})
and Crowther et al. (\cite{cro02}) estimated temperatures $T_{\star}$ of 
$\sim$85-100~kK. Following the models
of Smith et al. (\cite{smi}), this would not produce any detectable
\heii\ nebula since $Q({\rm He}^+)<2-5\times10^{40}$ photons s$^{-1}$
for $T_{\star}\le$100\,kK WC stars at $Z=0.2-0.4\,Z_{\odot}$. The non 
dectection and/or 
upper limits of the \heii\ $\lambda$4686 line around the aforementioned 
WC4 stars agree thus completely with the latest models. 

However the situation appears more complex for WN stars. The WN2 star
BAT99-5 has a spectrum almost identical to BAT99-2 (Foellmi et al. 
\cite{foelmc}), but no \heii\ emission was detected in its surroundings
($F$(\heii\ $\lambda$4686)/$F$(\hb)$<$0.04 in Pakull \cite{pak91b}). Hamann \&
Koesterke (\cite{ham00}) fitted the spectrum of BAT99-5 with a temperature 
of 71~kK, a value too low to produce enough He$^+$ ionizing photons. We 
note however that in the case of the Galactic WN2 star, WR2, Hamann \& Koesterke
(\cite{ham98}) found $T_{\star}=141\,{\rm kK}$, a temperature much larger 
than for BAT99-5. This is rather remarkable since the average temperature 
of all other WN subtypes is rather similar in the Galaxy and the LMC 
(Hamann \& Koesterke \cite{ham98}, Hamann \& Koesterke \cite{ham00},
Crowther et al. \cite{cro95a}, Crowther et al. \cite{cro95b}, Crowther 
\& Smith \cite{cro97}), though with a large scatter in both cases.
Logically, using all information available, both stars, BAT99-2 and 5, 
or none of them should exhibit nebular \heii\ emission.

Large discrepancies are also seen amongst WN4 stars. Hamann \& Koesterke 
(\cite{ham00}) found temperatures of 71 to 100~kK for the LMC WN4 stars :
the hottest of these stars should thus be capable of ionizing He$^+$.
For example, BAT99-94 (Brey 85) has a temperature of 100~kK, but yet
no nebular \heii\ was ever detected around this star (Niemela et al. \cite{nie91}),
contrary to BAT99-49. Moreover, the WN3 stars should in principle be
hotter than the WN4 stars, but no nebular \heii\ emission has been detected
around BAT99-62 and 82 ($F$(\heii\ $\lambda$4686)/$F$(\hb)$<$0.03 and 0.007, 
respectively, in Pakull \cite{pak91b}, Niemela et al. \cite{nie91}).

Finally, we note that in the remaining cases of \heii\ nebulae excited
by WR stars (WR102 in the Galaxy, IC1613\#3 in IC1613), these stars
present a WO spectral type. However, near the only WO star of the LMC, BAT99-123 
(Brey 93), no nebular \heii\ emission was detected ($F$(\heii\ 
$\lambda$4686)/$F$(\hb)$<$0.07 in Pakull \cite{pak91b}). Even in the case
of the largest concentration of hot stars in the LMC, i.e. in the 30 Doradus
complex, no nebular \heii\ emission was ever reported. 

BAT99-2, 49 and AB7 seem thus different from all other WR stars of the MCs,
but we find no particular reason for this uniqueness in the available data. 
Up to now, only 3 He$^+$ excitation mechanisms have been proposed: high 
velocity shocks - but we are not aware of any peculiar motions near these 
objects; X-ray ionization - but no bright X-ray source was ever recorded 
in the close neighbourhood of any of these stars; photoionization - but 
then why would these stars differ from (very) similar WN stars of the same 
Galaxy? Could binarity play a role? AB7 and BAT99-49 indeed have massive 
companions, but BAT99-2 appears single (Foellmi et al. \cite{foelmc}). 
Could they be mainly fossil \heii\ nebulae? Long-term monitoring of
both the nebulae and the stars would be required to unveil the signs of
recombination as well as the orbital motion due to the presence of
a putative compact companion in a wide eccentric orbit. 
However, up to now, only the visible spectrum of these 
peculiar stars is well studied, but the latter constitutes only a small part of
the hot stars' radiation. The UV and X-ray fluxes of the WR stars
of the MCs should be investigated deeply, since these peculiar stars 
may present particular features in these energy ranges.

\begin{sidewaystable*} 
\begin{center}
\caption{Same as Table \ref{lineratwr} for LMC BAT99-8, 11, 63 and 134. 
\label{lineratwr2}} 
\begin{tabular}{l | c c | c c | c | c c c c} 
\hline\hline
& \multicolumn{2}{c|}{BAT99-8$^a$}& \multicolumn{2}{c|}{BAT99-11}& BAT99-63& \multicolumn{4}{c}{BAT99-134}\\
 &  N&S& NE$^b$ & SW & NE& NE & near the star&SW &beyond SW$^c$\\ 
\hline
\oii\ 3727&  752(78) & 602 (63) & 449 (47) & 367 (38) & 220 (23) & 61 (6) & 238 (25) & 197 (21) & 463 (48)\\
H10&  && & 5.9 (0.6) &  &  &  &  & \\
H9&  & & & 10 (1) &  &  &  &  & \\
\neiii\ 3868&  12: & & & 4.2 (0.4) & 61 (6) & 106 (10) & 84 (8) & 78 (8) & 104 (10)\\
H8 + \hei& 15: && 17 (2) & 19 (2) & 19 (2) & 26 (3) & 23 (2) & 19 (2) & 27:\\
\neiii\ + \he & 24 (2) & & 13 (1) & 16 (2) & 27 (3) & 41 (4) & 37 (4) & 42 (4) & 40 (4)\\
\hd & 29 (3) & 24: & 27 (2) & 28 (3) & 33 (3) & 36 (3) & 32 (3) & 28 (3) & 30 (3)\\
\hg & 46 (4) & 41 (3) & 50 (4) & 51 (4) & 49 (4) & 51 (4) & 48 (4) & 51 (4) & 50 (4)\\
\oiii\ 4363& & & 4.9 (0.4) & 3.7 (0.3) & 5.1 (0.4) & 19 (2) & 12 (1) & 10 (1) & \\
\hei\ 4471&  & & 4.1 (0.3) & 4.1 (0.3) & 5.4 (0.4) &  & &3.9:  & \\
\hb &  100.& 100.& 100.& 100. & 100. & 100. & 100.& 100.& 100.\\
\oiii\ 4959& 61 (4) & 60 (4) & 78 (6) & 132 (9) & 237 (17) & 415 (29) & 284 (20) & 296 (21) & 256 (18)\\
\oiii\ 5007& 177 (13) & 184 (13) & 233 (17) & 393 (28) & 710 (51) & 1246 (89) & 846 (60) & 890 (63) & 770 (55)\\
\hei\ 5876 & 9.: & 12: & 12 (1) & 13 (1) & 13 (1) & 13 (1) & 15 (1) & 11 (1) & 12 (1)\\ 
\siii\ 6312&  & &  & 1.9 (0.2) & 2.3 (0.2) &  &  &  & \\
\nii\ 6548 & 21 (2) & 20 (2) & 7.8 (0.8) & 5.7 (0.6) & 6. (0.6) &  & 9.1 (0.9) & 5.1 (0.5) & 12 (1)\\
\ha&  286 (29) & 286 (29) & 279 (28) & 282 (28) & 286 (29) & 282 (28) & 282 (28) & 282 (28) & 282 (28)\\
\nii\ 6583& 52 (5) & 45 (5) & 26 (3) & 19 (2) & 19 (2) & 4.9: & 23 (2) & 17 (2) & 40 (4)\\
\hei\ 6678&  & & 2.7 (0.3) & 3.1 (0.3) & 2.7 (0.3) &  & 4.: & 3.6: & \\
\sii\ 6716& 87 (9) & 72 (7) & 20 (2) & 13 (1) & 24 (2) & 6.5: & 34 (4) & 25 (3) & 61 (6)\\
\sii\ 6731& 65 (7) & 56 (6) & 15 (2) & 9.4 (1.) & 17 (2) & 8.0: & 25 (3) & 18 (2) & 43 (4)\\
$F$(\hb) (10$^{-14}$ erg cm$^{-2}$ s$^{-1}$)& 0.85 & 0.44& 4.& 8.7& 4.1& 0.64& 1.2& 0.86& 1.\\
\hline
$A_{{\rm V}}$(mag)& 0.48 & 0.28& 0.59& 0.58& 0.35& 0.28& 0.22& 0.34& 0.34\\
$T_{{\rm e}}$\oiii\ (kK) &  & & 15.6$\pm$0.8 &11.4$\pm$0.4& 10.4$\pm$0.3 & 13.6$\pm$0.6 & 13.1$\pm$0.6& 12.1$\pm$0.5& \\
$n_{{\rm e}}$\sii\ (cm$^{-3}$) & $<$320 & $<$410& $<$380& $<$300& $<$280& & $<$290& $<$270& $<$260\\
\hline
He$^+$/H$^+\times 10^{2}$ & 6.5: & 8.6:& 8.3$\pm$0.4& 8.9$\pm$0.5 & 8.9$\pm$0.4 & 10.$\pm$0.9 & 11$\pm$0.5& 8.5$\pm$0.7& 8.8$\pm$0.8 \\
O$^+$/H$^+\times 10^{5}$ & 28$\pm$3 & 23$\pm$2& 3.4$\pm$0.4& 8.$\pm$0.8 & 7.$\pm$0.7 & 7.1$\pm$0.7 & 3.1$\pm$0.3& 3.5$\pm$0.4& 7.2$\pm$0.8 \\
O$^{2+}$/H$^+\times 10^{5}$ & 6.3$\pm$0.4 & 6.3$\pm$0.3& 2.2$\pm$0.1& 9.$\pm$0.4 & 22$\pm$1&17$\pm$1 & 13$\pm$1 & 17$\pm$1& 13$\pm$1 \\
$\rightarrow$ O/H$\times 10^{4}$ & 3.5$\pm$0.3 & 2.9$\pm$0.2& 0.56$\pm$0.04& 1.7$\pm$0.1 & 2.9$\pm$0.1 & 1.8$\pm$0.1 & 1.6$\pm$0.1& 2.$\pm$0.1& 2.$\pm$0.1 \\
N$^+$/H$^+\times 10^{6}$ & 11$\pm$1 & 10$\pm$1& 1.8$\pm$0.1& 2.5$\pm$0.2 & 3.2$\pm$0.2 & 0.5:& 2.6$\pm$0.2& 1.9$\pm$0.1& 4.4$\pm$0.3 \\
($\rightarrow$ N/O$\times 10^{2}$)$^d$ & 3.9$\pm$0.5 & 4.6$\pm$0.6& 5.3$\pm$0.7& 3.2$\pm$0.4 & 4.6$\pm$0.6 & 6.6:& 8.2$\pm$1.& 5.6$\pm$0.7& 6.1$\pm$0.8 \\
S$^+$/H$^+\times 10^{7}$ & 35$\pm$3 & 30$\pm$2& 3.2$\pm$0.2& 3.8$\pm$0.3 & 8.7$\pm$0.6 & 1.8:& 7.4$\pm$0.5& 6.4$\pm$0.5& 15$\pm$1 \\
S$^{2+}$/H$^+\times 10^{6}$ &  & & & 2.8$\pm$0.3 & 4.8$\pm$0.5 &  & & &  \\
Ne$^{2+}$/H$^+\times 10^{5}$ & 1.3: & & & 0.28$\pm$0.03 & 5.7$\pm$0.6 & 3.9$\pm$0.4 & 3.5$\pm$0.3& 4.2$\pm$0.4& 5.$\pm$0.5 \\
\hline
\end{tabular}
\end{center}
$^a$ Assuming $T_{{\rm e}}=10\,{\rm kK}$.\\
$^b$ Abundances shown below were derived for $T_{{\rm e}}=T$(\oiii)$=15.6\,{\rm kK}$.\\
$^c$ Assuming $T_{{\rm e}}=12.5\,{\rm kK}$ (close to the temperatures of the neighbouring regions).\\
$^d$ Assuming $N({\rm N}^+)/N({\rm O}^+)=N({\rm N})/N({\rm O})$.\\
\end{sidewaystable*}

\section{Conclusion}

In this paper and a previous letter (Paper I), we investigate the 
relations between hot stars of the MCs and their environment. We first 
focus on the peculiar \heii\ nebulae ionized by WR stars. We provide 
here the first high quality \heii\ $\lambda$4686
images of these high excitation nebulae. The \heii\ nebula around 
BAT99-49 appears more extended but also fainter than the one associated 
with BAT99-2. The high excitation nebula inside N76 occupies a size 
of roughly half the one of the \ha\ nebula. It is centered on the WR 
star AB7. Once calibrated, the images enable us to estimate the He$^+$ 
ionizing fluxes. For BAT99-2 and 49, we found $Q({\rm He}^+)$ of 4 and 
1$\times$10$^{47}$ photons s$^{-1}$, respectively. In the latest 
theoretical models, this corresponds to 90-100~kK WN stars. AB7 
apparently emits $Q({\rm He}^+)=5\times10^{48}$ photons s$^{-1}$,
and no recent model could explain such a high value. 
In addition, after comparing these objects to other WR stars of the 
MCs, we find no particular reason for their uniqueness: other very
similar stars do not present such high excitation features. 

In this study, we also used FORS spectroscopy to search for \heii\ emission
around seven other WR stars: BAT99-8, 9, 11, 52, 63, 84 and 134. We do not 
find any new \heii\ nebula, even near BAT99-9 where hints of nebular \heii\ 
emission had been found in a previous study. These data enable us
to analyse the physical properties of the above nebulae and also of the ring 
nebulae associated with BAT99-8, 11, 63 and 134. The abundance determinations
show only small chemical enrichment for the nebulae around BAT99-2, 49, 
100 and AB7. The slightly larger N/O and the slightly lower oxygen abundance
are incompatible with a single circumstellar bubble, but could be explained
in a scenario where the circumstellar bubble blown by the WR star is currently 
merging with the interstellar bubble blown by its progenitor. In contrast, 
N44C and the nebulae associated with BAT99-8 and 63 present normal abundances,
suggesting ISM dominated bubbles.

We also examine a fourth \heii\ nebula of the MCs, N44C. It is thought to be
a fossil X-ray nebula in which \heii\ should recombine with timescales
of 20-100 yr. Using our VLT data, we estimate $L$(\hb)$=9\times10^{36}$ 
erg s$^{-1}$ and $L$(\heii\ $\lambda$ 4686)$=4\times10^{35}$ erg s$^{-1}$.
 Comparing our data
with older values, we find no sign of variation in the \heii\ $\lambda$4686
line strength. This puzzling result indicates a much longer recombination
timescale, and thus a very low density. Alternatively, another, unknown, 
ionizing mechanism could be at work. The X-ray source may also have 
turned on again some time in between the previous observations and ours. Further 
monitoring is recommended to follow the possible evolution of the \heii\ and 
X-ray emissions, and also to search for traces of orbital motion in Star \#2, 
and thus of a companion that wculd be responsible for the episodic X-ray 
emission as required by the fossil X-ray nebula scenario.

\begin{acknowledgements}
We acknowledge support from the PRODEX XMM-OM and Integral Projects 
and through contracts P4/05 and P5/36 `P\^ole d'attraction Interuniversitaire' 
(Belgium). We thank the referee, Dr M. Heydari-Malayeri, for his useful suggestions.
\end{acknowledgements}


\begin{thebibliography}{}

\bibitem[2001]{bar} Bartzakos, P., Moffat, A.F.J., \& Niemela, V.S. 2001, MNRAS, 324, 18
\bibitem[1999]{ben} Benjamin, R.A., Skillman, E.D., \& Smits, D.P. 1999, ApJ, 514, 307
\bibitem[1999]{bic} Bica, E.L.D., Schmitt, H.R., Dutra, C.M., \& Oliveira, H.L. 1999, AJ, 117, 238
\bibitem[1985]{bou} Bouchet, P., Lequeux, J., Maurice, E., Pr\'evot, L., \& Pr\'evot-Burnichon, M.L. 1985, A\&A, 149, 330
\bibitem[1985]{cap} Caplan, J., \& Deharveng, L. 1985, A\&AS, 62, 63
\bibitem[1989]{car} Cardelli, J.A., Clayton, G.C., \& Mathis, J.S. 1989, ApJ, 345, 245
\bibitem[1981]{chu81} Chu, Y.-H. 1981, PhD thesis, University of California, Berkeley
\bibitem[1983]{chu83} Chu, Y.-H. 1983, ApJ, 269, 202
\bibitem[1999]{chu99} Chu, Y.-H., Weis, K., \& Garnett, D.R. 1999, AJ, 117, 1433
\bibitem[1995a]{cro95a} Crowther, P.A., Hillier, D.J., \& Smith, L.J. 1995a, A\&A, 293, 403
\bibitem[1995b]{cro95b} Crowther, P.A., Hillier, D.J., \& Smith, L.J. 1995b, A\&A, 302, 457
\bibitem[1997]{cro97} Crowther, P.A., \& Smith, L.J. 1997, A\&A, 320, 500
\bibitem[2002]{cro02} Crowther, P.A., Dessart, L., Hillier, D.J., Abbott, J.B., \& Fullerton, A.W. 2002, A\&A, 392, 653
\bibitem[1994]{dop} Dopita, M.A., Bell, J.F., Chu, Y.-H., \& Lozinskaya, T.A. 1994, ApJS, 93, 455
\bibitem[1997]{dop97} Dopita, M.A., \& Hua, C.T. 1997, ApJS, 108, 515
\bibitem[1975]{duf} Dufour, R.J. 1975, ApJ, 195, 315
\bibitem[1989]{fit} Fitzpatrick, E.L. 1989, IAUS, 135, 37
\bibitem[2003a]{foesmc} Foellmi, C., Moffat, A.F.J., \& Guerrero, M.A. 2003a, MNRAS, 338, 360
\bibitem[2003b]{foelmc} Foellmi, C., Moffat, A.F.J., \& Guerrero, M.A. 2003b, MNRAS, 338, 1025
\bibitem[1996]{gar} Garc\'{\i}a-Segura, G., Langer, N., \& MacLow, M.-M. 1996, A\&A, 316, 133
\bibitem[1991a]{gar91a} Garnett, D.R., Kennicutt, R.C.Jr., Chu, Y.-H., \& Skillman, E.D. 1991a, PASP, 103, 850
\bibitem[1991b]{gar91b} Garnett, D.R., Kennicutt, R.C.Jr., Chu, Y.-H., \& Skillman, E.D. 1991b, ApJ, 373, 458
\bibitem[2000]{gar00} Garnett, D.R., Galarza, V.C., \& Chu, Y.-H. 2000, ApJ, 545, 251
\bibitem[1984]{gou} Goudis, C., \& Meaburn, J. 1984, A\&A, 137, 152
\bibitem[1998]{gra} Gr\"afener, G., Hamann, W.-R., Hillier, D.J., \& Koesterke, L. 1998, A\&A, 329, 190
\bibitem[1998]{ham98} Hamann, W.-R., \& Koesterke, L. 1998, A\&A, 333, 251
\bibitem[2000]{ham00} Hamann, W.-R., \& Koesterke, L. 2000, A\&A, 360, 647
\bibitem[2001]{hey} Heydari-Malayeri, M., Charmandaris, V., Deharveng, L., Rosa, M.R., Schaerer, D., \& Zinnecker, H. 2001, A\&A, 372, 527
\bibitem[2000]{mas} Massey, P., Waterhouse, E., \& Degioia-Eastwood, K. 2000, AJ, 119, 2214
\bibitem[1985]{mcc} McCall, M.L., Rybski, P.M., \& Shields, G.A. 1985, ApJS, 57, 1 
\bibitem[1991]{mel} Melnick, J., \& Heydari-Malayeri, M. 1991, IAUS, 143, 409
\bibitem[2001]{naz01} Naz\'e, Y., Chu, Y.-H., Points, S.D., Danforth, C.W., Rosado, M.,
\& Chen, C.-H.R. 2001, AJ, 122, 921
\bibitem[2002]{naz02} Naz\'e, Y., Chu, Y.-H., Guerrero, M.A., Oey, M.S., Gruendl, R.A., \& Smith, R.C. 2002, AJ, 124, 3325
\bibitem[2003]{naz03} Naz\'e, Y., Rauw, G., Manfroid, J., Chu, Y.-H., \& Vreux, J.-M. 2003, A\&A, 401, L13 (paper I)
\bibitem[1991]{nie91} Niemela, V.S., Heathcote, S.R., \& Weller, W.G. 1991, IAUS, 143, 425
\bibitem[2002]{nie02} Niemela, V.S., Massey, P., Testor, G., \& Gim\'enez-Ben\'{\i}tez, S. 2002, MNRAS, 333, 347
\bibitem[1995]{oey} Oey, M.S., \& Massey, P. 1995, ApJ, 452, 210
\bibitem[1990]{oke} Oke, J.B. 1990, AJ, 99, 1621
\bibitem[1989a]{pak89a} Pakull, M.W., \& Motch, C. 1989a, Nature, 337, 337
\bibitem[1989b]{pak89b} Pakull, M.W., \& Motch, C. 1989b, Extranuclear Activity in Galaxies, eds. Meurs \& Fosbury (Garching bei Munchen), 285
\bibitem[1991]{pak91a} Pakull, M.W., \& Bianchi, L. 1991, IAUS, 143, 260
\bibitem[1991]{pak91b} Pakull, M.W. 1991, IAUS 143, 391
\bibitem[1990]{rus} Russell, S.C., \& Dopita, M.A. 1990, ApJS, 74, 93
\bibitem[1996]{sch} Schaerer D. 1996, ApJ, 467, L 17
\bibitem[1995]{sha} Shaw, R.A., \& Dufour, R.J.  1995, PASP, 107, 896
\bibitem[2002]{smi} Smith, L.J., Norris, R.P.F.,\& Crowther, P.A. 2002, MNRAS, 337, 1309
\bibitem[1986]{sta} Stasi\'nska, G., Testor, G., \& Heydari-Malayeri, M. 1986, A\&A, 170, L4
\bibitem[1995]{sto} Storey, P.J., \& Hummer, D.G. 1995, MNRAS, 272, 41
\bibitem[1983]{tuo} Tuohy, I.R., \& Dopita, M.A. 1983, ApJ, 268, L11
\bibitem[2002]{wal} Walborn, N.R. 2002, Hot Star Workshop III: The Earliest Stages of Massive Star Birth. ASP Conf. Proc. Vol. 267, ed. P.A. Crowther (San Francisco), 111
\end{thebibliography}
\end{document}